\documentclass[twoside]{article}
\usepackage[pdftex]{graphicx}
\catcode`\@=11 \long\def\@makefntext#1{ \protect\noindent \hbox to
3.2pt {\hskip-.9pt
$^{{\eightrm\@thefnmark}}$\hfil}#1\hfill}       

\def\@makefnmark{\hbox to 0pt{$^{\@thefnmark}$\hss}}    

\def\ps@myheadings{\let\@mkboth\@gobbletwo
\def\@oddhead{\hbox{}
\rightmark\hfil\eightrm\thepage}
\def\@oddfoot{}\def\@evenhead{\eightrm\thepage\hfil
\leftmark\hbox{}}\def\@evenfoot{}
\def\sectionmark##1{}\def\subsectionmark##1{}}

\oddsidemargin=\evensidemargin 
\newcounter{sectionc}\newcounter{subsectionc}\newcounter{subsubsectionc}
\renewcommand{\section}[1] {\vspace{12pt}\addtocounter{sectionc}{1}
\setcounter{subsectionc}{0}\setcounter{subsubsectionc}{0}\noindent
    {\tenbf\thesectionc. #1}\par\vspace{5pt}}
\renewcommand{\subsection}[1] {\vspace{12pt}\addtocounter{subsectionc}{1}
    \setcounter{subsubsectionc}{0}\noindent
    {\bf\thesectionc.\thesubsectionc. {\kern1pt \bfit #1}}\par\vspace{5pt}}
\renewcommand{\subsubsection}[1] {\vspace{12pt}\addtocounter{subsubsectionc}{1}
    \noindent{\tenrm\thesectionc.\thesubsectionc.\thesubsubsectionc.
    {\kern1pt \tenit #1}}\par\vspace{5pt}}
\newcommand{\nonumsection}[1] {\vspace{12pt}\noindent{\tenbf #1}
    \par\vspace{5pt}}

\newcounter{appendixc}
\newcounter{subappendixc}[appendixc]
\newcounter{subsubappendixc}[subappendixc]
\renewcommand{\thesubappendixc}{\Alph{appendixc}.\arabic{subappendixc}}
\renewcommand{\thesubsubappendixc}
    {\Alph{appendixc}.\arabic{subappendixc}.\arabic{subsubappendixc}}

\renewcommand{\appendix}[1] {\vspace{12pt}
        \refstepcounter{appendixc}
        \setcounter{figure}{0}
        \setcounter{table}{0}
        \setcounter{lemma}{0}
        \setcounter{theorem}{0}
        \setcounter{corollary}{0}
        \setcounter{definition}{0}
        \setcounter{equation}{0}
        \renewcommand{\thefigure}{\Alph{appendixc}.\arabic{figure}}
        \renewcommand{\thetable}{\Alph{appendixc}.\arabic{table}}
        \renewcommand{\theappendixc}{\Alph{appendixc}}
        \renewcommand{\thelemma}{\Alph{appendixc}.\arabic{lemma}}
        \renewcommand{\thetheorem}{\Alph{appendixc}.\arabic{theorem}}
        \renewcommand{\thedefinition}{\Alph{appendixc}.\arabic{definition}}
        \renewcommand{\thecorollary}{\Alph{appendixc}.\arabic{corollary}}
        \renewcommand{\theequation}{\Alph{appendixc}.\arabic{equation}}
        \noindent{\tenbf Appendix \theappendixc #1}\par\vspace{5pt}}
\newcommand{\subappendix}[1] {\vspace{12pt}
        \refstepcounter{subappendixc}
        \noindent{\bf Appendix \thesubappendixc. {\kern1pt \bfit #1}}
    \par\vspace{5pt}}
\newcommand{\subsubappendix}[1] {\vspace{12pt}
        \refstepcounter{subsubappendixc}
        \noindent{\rm Appendix \thesubsubappendixc. {\kern1pt \tenit #1}}
    \par\vspace{5pt}}

\newcommand{\textlineskip}{\baselineskip=13pt}
\newcommand{\smalllineskip}{\baselineskip=10pt}

\def\eightcirc{
\begin{picture}(0,0)
\put(4.4,1.8){\circle{6.5}}
\end{picture}}
\def\eightcopyright{\eightcirc\kern2.7pt\hbox{\eightrm c}}

\newcommand{\copyrightheading}[1]
    {\vspace*{-2.5cm}\smalllineskip{\flushleft
    {\footnotesize International Journal of Modern Physics D, in press #1}\\
    {\footnotesize World Scientific Publishing Company, 2005}\\
     }}

\newcommand{\publisher}[2]{{\begin{center}\footnotesize\smalllineskip
    Received #1\\
    Revised #2
    \end{center}
    }}

\def\abstracts#1#2#3{{
    \centering{\begin{minipage}{4.5in}\footnotesize\baselineskip=10pt
    \parindent=0pt #1\par
    \parindent=15pt #2\par
    \parindent=15pt #3
    \end{minipage}}\par}}

\renewenvironment{thebibliography}[1]
    {\frenchspacing
     \ninerm\baselineskip=11pt
     \begin{list}{\arabic{enumi}.}
    {\usecounter{enumi}\setlength{\parsep}{0pt}
     \setlength{\leftmargin 12.7pt}{\rightmargin 0pt} 
     \setlength{\itemsep}{0pt} \settowidth
    {\labelwidth}{#1.}\sloppy}}{\end{list}}

\newcounter{itemlistc}
\newcounter{romanlistc}
\newcounter{alphlistc}
\newcounter{arabiclistc}

\newcommand{\fcaption}[1]{
        \refstepcounter{figure}
        \setbox\@tempboxa = \hbox{\footnotesize Fig.~\thefigure. #1}
        \ifdim \wd\@tempboxa > 5in
           {\begin{center}
        \parbox{5in}{\footnotesize\smalllineskip Fig.~\thefigure. #1}
            \end{center}}
        \else
             {\begin{center}
             {\footnotesize Fig.~\thefigure. #1}
              \end{center}}
        \fi}

\newcommand{\tcaption}[1]{
        \refstepcounter{table}
        \setbox\@tempboxa = \hbox{\footnotesize Table~\thetable. #1}
        \ifdim \wd\@tempboxa > 5in
           {\begin{center}
        \parbox{5in}{\footnotesize\smalllineskip Table~\thetable. #1}
            \end{center}}
        \else
             {\begin{center}
             {\footnotesize Table~\thetable. #1}
              \end{center}}
        \fi}

\def\@citex[#1]#2{\if@filesw\immediate\write\@auxout
    {\string\citation{#2}}\fi
\def\@citea{}\@cite{\@for\@citeb:=#2\do
    {\@citea\def\@citea{,}\@ifundefined
    {b@\@citeb}{{\bf ?}\@warning
    {Citation `\@citeb' on page \thepage \space undefined}}
    {\csname b@\@citeb\endcsname}}}{#1}}

\newif\if@cghi
\def\cite{\@cghitrue\@ifnextchar [{\@tempswatrue
    \@citex}{\@tempswafalse\@citex[]}}
\def\citelow{\@cghifalse\@ifnextchar [{\@tempswatrue
    \@citex}{\@tempswafalse\@citex[]}}
\def\@cite#1#2{{$\null^{#1}$\if@tempswa\typeout
    {IJCGA warning: optional citation argument
    ignored: `#2'} \fi}}

\def\pmb#1{\setbox0=\hbox{#1}
    \kern-.025em\copy0\kern-\wd0
    \kern.05em\copy0\kern-\wd0
    \kern-.025em\raise.0433em\box0}

\def\fnt#1#2{\footnotetext{\kern-.3em
    {$^{\mbox{\scriptsize #1}}$}{#2}}}

\def\@makefnmark{\hbox to 0pt{$^{\@thefnmark}$\hss}}    

\def\ps@myheadings{%
    \let\@oddfoot\@empty\let\@evenfoot\@empty
    \def\@evenhead{\slshape\leftmark\hfil}
    \def\@oddhead{\hfil{\slshape\rightmark}}
    \let\@mkboth\@gobbletwo
    \let\sectionmark\@gobble
    \let\subsectionmark\@gobble
    }
\font\tenrm=cmr10 \font\tenit=cmti10 \font\tenbf=cmbx10
\font\bfit=cmbxti10 at 10pt \font\ninerm=cmr9 
 \font\eightrm=cmr8

\def\qed{\hbox{${\vcenter{\vbox{            
   \hrule height 0.4pt\hbox{\vrule width 0.4pt height 6pt
   \kern5pt\vrule width 0.4pt}\hrule height 0.4pt}}}$}}

\begin{document}
\setlength{\textheight}{7.7truein}  

\thispagestyle{empty}

\markboth{\protect{\footnotesize\it Constraint on
intermediate-range gravity from earth-satellite and lunar orbiter
measurements, and lunar laser ranging}}{\protect{\footnotesize\it
Guangyu Li \& Haibin Zhao}}

\normalsize\textlineskip

\setcounter{page}{1}

\copyrightheading{} 

\vspace*{0.88truein}

\centerline{\bf Constraint on intermediate-range gravity from
earth-satellite and lunar orbiter } \vspace*{6pt} \centerline{\bf
measurements, and lunar laser ranging}

\vspace*{0.37truein}

\vspace*{10pt} \centerline{\footnotesize Guangyu Li and Haibin
Zhao\footnote{Email: gyl@pmo.ac.cn} }\baselineskip=12pt
\centerline{\footnotesize\it Purple Mountain Observatory, Chinese
Academy of Sciences} \baselineskip=10pt
\centerline{\footnotesize\it Nanjing 210008, P.R.China}
\vspace*{0.005truein} \baselineskip=12pt
\centerline{\footnotesize\it National Astronomical Observatories,
Chinese Academy of Sciences} \baselineskip=10pt
\centerline{\footnotesize\it Beijing 100012, P.R.China}
\vspace*{0.225truein}

\publisher{December 23, 2004}{March 10, 2005}

\vspace*{0.21truein} \abstracts{In the experimental tests of
gravity, there have been considerable interests in the possibility
of intermediate-range gravity. In this paper, we use the
earth-satellite measurement of earth gravity, the lunar orbiter
measurement of lunar gravity, and lunar laser ranging measurement
to constrain the intermediate-range gravity from $\lambda=1.2
\times 10^{7} \rm\: m-3.8 \times 10^{8} \rm\: m$. The limits for
this range are $\alpha=10^{-8}- 5 \times 10^{-8}$, which improve
previous limits by about one order of magnitude in the range
$\lambda=1.2 \times 10^{7} \rm\: m-3.8 \times 10^{8} \rm\:
m$.}{}{}


\vspace*{1pt}\textlineskip  
\section{Introduction}    
\vspace*{-0.5pt}

\noindent To test theories of gravity and to probe its origin,
there have been considerable interests in the search for
non-Newtonian gravity.\cite{1} Both composition-dependent and
composition-independent deviations from Newtonian gravity have
been searched for. Composition-independent deviations from the
Newton's inverse square law between two point masses $M$ and $m$
can be expressed in terms of a distance-dependent gravitational
``constant" $G(r)$ as

\begin{equation}\label{1}
  G(r)=F_{grav} r^{2} / Mm,
\end{equation}

\noindent where $r$ is the separation distance between two masses.
Wagonar,\cite{2} Fujii,\cite{3} and O'Hanlon\cite{4} have provided
a theoretical basis for this variation and proposed $\alpha-\mu$
model as

\begin{equation}\label{2}
  G(r)=G_{c} [ 1 + \alpha(1+\mu r) e^{-\mu r}],
\end{equation}

\noindent where $G_{c}$, $\alpha$, and $\mu(=\lambda^{-1})$ are
constants, and $G(r)$ as a function of $r$ is shown in Fig. 1. As
shown in the figure, $G(r)$ goes to $G_{c}[1+\alpha]$ with
decreasing distance , and goes to $G_{c}$ with increasing
distance.



Since the proposal of the $\alpha-\mu$ model, there have been
theoretical analyses and experimental efforts to constrain the
variable ranges of $\alpha$ and $\mu$.\cite{1,5,6,7} The cumulated
constraints together with the result obtained in this paper are
shown in Fig. 2.

In this paper, we use the satellite measurement of earth gravity,
the lunar orbiter measurement of lunar gravity, and lunar laser
ranging measurement to constrain the intermediate-range gravity
from $\lambda=1.2 \times 10^{7} \rm\: m-3.8 \times 10^{8} \rm\:
m$. In section 2, we discuss the value of the $GM_{earth}$ from
satellite measurement of earth gravity. In section 3, we discuss
the value of $GM_{moon}$ from the lunar orbiter measurement of
lunar gravity. In section 4, we discuss the $G'M_{earth+moon}$
obtained from the lunar laser ranging experiment. In section 5, we
fit the $\alpha$ parameter of the $\alpha-\mu$ model to these
values to constrain the intermediate-range gravity. In section 6,
we look into possible improvements in the future.

\section{Measurement of $GM_{earth}$ from Satellite Orbit Determination}

\noindent The geocentric gravitational coefficient of the earth
($GM_{earth}$) is a key parameter of earth gravity model. It is
determined by observing the influence of this parameter on the
motion of earth satellites. LAGEOS (LAser GEOdynamics Satellite),
launched in 1976, was designed to minimize the effects of
non-gravitational forces. Because of its high altitude (5900 km),
the high-{\it l} components of the earth's geopotential are
greatly attenuated. The result of $GM_{earth}$ determined by
LAGEOS satellite has been improved consistently.\cite{8} In 1985,
a value of $398600.440 \pm 0.002 \rm\: km^{3}/sec^{2}$ for
$GM_{earth}$ was determined based on 8 years data of laser ranging
to LAGEOS by Tapley {\it et al.}\cite{9} This improved by one
order of magnitude than that of $GM_{earth}$ ($398600.44 \pm 0.02
\rm\: km^{3}/sec^{2}$)\cite{10} determined using laser ranging to
4 near-earth satellites in 1978. In 1989, Ries {\it et
al.}\cite{11} reported a solution for $GM_{earth}$ obtained from
LAGEOS laser ranging and also from a multi-satellite solution to
be $398600.4405 \pm 0.001 \rm\: km^{3}/sec^{2}$.

In 1992, Ries {\it et al.}\cite{8} found that the value of the
correction for the offset between the LAGEOS center-of-mass and
the effective reflecting surface should be shifted by 11 mm. After
the center-of-mass offset error had been corrected, they improved
determination of $GM_{earth}$ by two methods. The first method is
the LAGEOS-only solution, where the data set consisted of laser
ranges from over 60 stations spanning the five-year period from
November 1986 to November 1991. The second one is that laser
range, Doppler and altimeter observations from 17 near-earth
satellites are combined with surface gravity data in a solution
for $GM_{earth}$. The effects of general relativity are taken into
account in these two methods. The determined values for
$GM_{earth}$ by two above-mentioned methods are remarkable in
agreement with each other. The value of $GM_{earth}$ determined
there is $398600.4418 \pm 0.0008 \rm\: km^{3}/sec^{2}$ in SI units
or $398600.4415 \pm 0.0008 \rm\: km^{3}/sec^{2}$ in TT
(Terrestrial Time) units. The scale difference of $L_{G} =
6.969290134 \times 10^{-10}$ between the TT and SI units results
in a change of $\Delta GM_{earth} = 0.0003\rm\:
km^{3}/sec^{2}$.\cite{11,12} The value of $GM_{earth}$, including
the mass of the earth's atmosphere, has been estimated with
$1$-$\sigma$ uncertainties in the measurement data. The $rms$ of
the residuals from the short-arc fit was 2.8 cm.\cite{8} UT/CSR
(University of Texas/Center of Space Research) has obtained a
solution of the gravity field complete to degree and order 50 by
the second method. Since then more endeavors have been taken into
developing earth gravity model with higher degree and order, while
the value of $GM_{earth}$ has not been improved. In fact, the
value is still adopted by the IERS(2003) standard released
recently.\cite{13}

A new EGM96 gravity Model, which was proposed by NASA (National
Aeronautics and Space Administration) , NIMA (National Imagery and
Mapping Agency), and OSU (Ohio State University ) in 1993, is
developed to support terrestrial and extraterrestrial scientific
endeavors in connection with its associated global geoid. That is
one of most accurate earth gravity models and its geoid undulation
with respect to the WGS84 (World Geodetic System 1984) ellipsoid
is about $¡À0.5 \rm\: m$ to $¡À1.0 \rm\: m$.

CHAMP\cite{14} (CHAllenging Minisatellite Payload), which was
launched on July 15, 2000, was proposed by DLR (Deutschen Zentrum
f\"{u}r Luft- und Raumfahrt) in 1994 and managed by GFZ
(GeoForschungsZentrum Potsdam). Determining the earth gravity
field is one of main scientific objectives of this project. The
CHAMP project improves by two orders of magnitude in precision for
determining the earth gravity field. Thus CHAMP obtains low
frequency characteristics of high-precision earth static gravity
field model and their variations with time.
EIGEN-CHAMP03S\cite{15} is a CHAMP-only gravity field model
derived from CHAMP GPS satellite-to-satellite and accelerometer
data out of the period from October 2000 to June 2003. The
accuracy of EIGEN-CHAMP03S is about $5 \rm\: cm$ and $0.5 \rm\:
mgal$ in terms of geoid heights and gravity anomalies,
respectively, with space resolution at $\lambda/2=400 \rm\: km$.

GRACE,\cite{16} which was launched on March 17, 2002, is a joint
project between NASA and DLR. The primary objective of the GRACE
mission is to provide with unprecedented accuracy estimates of the
global high-resolution models of the earth's gravity field for a
period of up to five years. With the increase of GRACE measurement
data, several Gravity field models derived by GRACE-only data or
by combined GRACE and CHAMP data have been developed, including
GGM01S (complete to degree/order 120), GRACE02S (complete to
degree/order 150), and GGM02S (complete to degree/order
200).\cite{17}

The value of $GM_{earth}$ accepted by all of above-mentioned earth
gravity models [IERS(2003), EGM96, EIGEN-CHAMP03S, GGM01S,
GRACE02S, GGM02S] is that one determined by Ries {\it et
al.}\cite{8} in 1992, i.e.,

\begin{equation}\label{3}
GM_{earth} = 398600.4418 \pm 0.0008 \rm\: km^{3}/sec^{2}. \qquad
(SI\; units)
\end{equation}

\section{Measurement of $GM_{moon}$ from the Orbit Determination of Lunar Orbiters}

\noindent Like earth, the $GM_{moon}$ of the Moon is also a key
parameter of Moon gravity field model. In early period, Lunar
Orbiter series and Apollo series space projects from later 1960s
to early 1970s have obtained large numbers of data for developing
Moon gravity field model. The Clementine mission was launched on
January 25, 1994. One of its key objectives is to establish a
high-resolution gravity field model of the Moon by measurements of
perturbations in the motion of the spacecraft to infer the lunar
gravity field. The GLGM-2 (Goddard Lunar Gravity Model 2)\cite{18}
is the final solution of the Clementine gravity model. The gravity
solution is based on Doppler tracking of Lunar Orbiters 1 to 5,
the Apollo-15 and Apollo-16 subsatellites, and the Clementine
spacecraft.\cite{18} In order to obtain a more realistic estimate
of the uncertainty of gravity field, the solution was calibrated
by the use of subset solutions, as described by Lerch.\cite{19}
The value of $GM_{moon}$ for the GLGM-2 is $4902.80295 \pm 0.00224
\rm\: km^{3}/sec^{2}$.

On January 6, 1998, NASA launched Lunar Prospector spacecraft,
which is the third Discovery series mission. For about one-year,
Lunar Prospector was placed in a near circular orbit at an
altitude 100 km and remained in this 2-hour orbit with
$90^{\circ}$ inclination. This provided global coverage for the
lunar gravity experiment every 14 days. On December 19, 1998, the
altitude of spacecraft was reduced to an average of 40 km to
calibrate the gravity field in preparation for an even lower
extended mission. After January 29, 1999, Lunar Prospector began
its extended mission when the spacecraft was lowered to an average
of 30 km to obtain higher resolution gravity data.\cite{20}

Since there is no direct measurement of the lunar farside from
Lunar Prospector or any other mission, gravity details for the
farside of the moon are quite limited. However, it has almost no
influence on the value of $GM_{moon}$. The 100th-degree lunar
gravity models (LP100J and LP100K) extract most of the information
from the nominal 100 km altitude.\cite{20} In these Lunar
Prospector models series, the values of $GM_{moon}$ are listed in
Table 1, where the value of LP100K model, $4902.800238 \pm
0.000206 \rm\: km^{3}/sec^{2}$, is the most accurate one, i.e.,

\begin{equation}\label{4}
GM_{moon}=4902.800238 \pm 0.000206 \rm\: km^{3} / sec^{2}.  \qquad
(SI\; units)
\end{equation}

\begin{table}[htbp]
\tcaption{ Key parameters of Lunar Prospector series gravity field
models.\cite{21}} \centerline{\footnotesize\smalllineskip
\begin{tabular}{l l l l l l l l}\\
\hline \hline
    &{Model}  &{$GM_{moon}$}            &{Degree/order}             \\
    &{     }  &{($\rm km^{3}/sec^{2}$)} &{ }                        \\
\hline \hline
    &{LP75D}  &{$4902.801374 \pm 0.00031$}   &{75}                  \\
    &{LP75G}  &{$4902.800269 \pm 0.000233$}  &{75}                  \\
    &{LP100J} &{$4902.800476 \pm 0.000209$}  &{100}                 \\
    &{LP100K} &{$4902.800238 \pm 0.000206$}  &{100}                 \\
\hline\\
\end{tabular}}
\end{table}

\section{Determination of $G'M_{earth+moon}$ from the Lunar Laser Ranging Experiment}

The accuracy of Lunar Laser Ranging have improved during last 35
years to 2 cm. The accuracy of Lunar Laser Ranging data from 1970
to 1997 adopted by DE 405, is about 2-3 cm; the relative accuracy
of ranging is better than $8 \times 10^{-11}$. The latest DE 410
has been created especially for the positions of Mars and Saturn,
and it includes recent determinations of $GM$ values of the earth
and the moon. In fact, the $GM$ values discussed in section 2 and
3 are adopted by DE 410.\cite{22} In the earth-moon center-of-mass
reference frame, the Newtonian equation of the motion of moon is

\begin{equation}\label{5}
  \ddot{\vec{r}}=- \frac{G'M_{earth+moon}}{r^{3}} \vec{r},
\end{equation}

\noindent where $G'$ is the gravitational constant at earth-moon
distance, i.e., 386000 km.

In DE 405, the GM constant of earth-moon system, which is in TDB
(Barycentric Dynamical Time) Scale,\cite{23} derived using Lunar
Laser Ranging data is \cite{13}

\begin{eqnarray}\label{6}
  G'M_{earth+moon} & = & 0.8997011347 \times 10^{-9} \rm\:  AU^{3}/day^{3}  \nonumber \\
                   & = & 403503.23348 \rm\: km^{3} / sec^{2}.  \qquad (TDB\; scale)
\end{eqnarray}

\noindent When this is converted to SI units, it should be
multiplied by a factor of $1 + L_{B}$ with $L_{B}=1.55051976772
\times 10^{-8} \pm (2 \times 10^{-17})$,\cite{11,12} and the value
is $403503.23974 \rm\; km^{3} / sec^{2}$.

\noindent The uncertainty of this value is not given in the
literature of DE 405. However, we can estimate the uncertainty
conservatively using the accuracy of Lunar Laser Ranging data
explained at the beginning of this section, i.e., $8 \times
10^{-11}$ in fraction of $r$; for $\frac{\delta(G'M)}{G'M} =
2\times \frac{\delta r}{r}$, the fractional uncertainty of
$G'M_{earth+moon}$ is $1.6 \times 10^{-11}$ or $\pm 0.00006 \rm\:
km^{3} / sec^{2}$. Putting the value and uncertainty together,
$G'M_{earth+moon}$ is

\begin{equation}\label{7}
  G'M_{earth+moon}=403503.23974 \pm 0.00006 \rm\: km^{3} / sec^{2}.
   \qquad  (SI\; units)
\end{equation}

\section{Constraint on the $\alpha-\mu$ model}

\noindent Comparing with the separation distance from earth to
moon, the distance of $(1.8-12) \times 10^{6} \rm\: m$, on which
the gravity field models based on earth's satellite or moon's
satellite are constructed, is rather short. For these short
distances, we can obtain approximatively

\begin{equation}\label{8}
  M_{moon} / M_{earth}\simeq 0.0123000371 \pm (5 \times 10^{-10}),
\end{equation}

\noindent from (3), (4) in section 2 and 3. DE 410 also gives this
mass ratio as $0.0123000371$; the same as (8) without quoting an
uncertainty. However, DE 405 adopts the value as $0.0123000383$.

From (7) and (8), we obtain

\begin{equation}\label{9}
  G'M_{earth} = 398600.4395 \pm 0.00022 \rm\: km^{3} / sec^{2}.
\end{equation}

In (9), $G'$ is measured at the separation distance from earth to
moon of $r_{m}=3.86 \times 10^{8} \rm\: m$. According to section
2, at the distance of earth satellite of $r_{s}=1.2 \times 10^{7}
\rm\: m$, we have

\begin{equation}\label{10}
  GM_{earth}=398600.4418 \pm 0.0008 \rm\: km^{3} / sec^{2}.
\end{equation}

Dividing (10) by (9), we obtain

\begin{equation}\label{11}
  G / G'=1.000000006 (\pm 6\times 10^{-9}). \qquad  (3-\sigma)
\end{equation}

According to $\alpha-\mu$ model, now we use (11) to constrain the
parameter space of intermediate range gravity (2). Assuming $r_{s}
< \lambda <r_{m}$ and setting

\begin{equation}\label{12}
  \Psi(\mu,r)=(1+\mu r)e^{-\mu r},\qquad  (\mu = \lambda^{-1})
\end{equation}

\noindent (11) becomes

\begin{equation}\label{13}
  \frac{1+\alpha\Psi(\mu,r_{s})}{1+\alpha\Psi(\mu,r_{m})}-1=(6 \pm 6) \times
  10^{9}.
\end{equation}

In (13), $\alpha$ is small and (13) can be approximated as

\begin{equation}\label{14}
  [ \Psi (\mu, r_{s})- \Psi (\mu, r_{m})] \alpha = (6 \pm 6) \times 10^{-9}.
\end{equation}

\noindent Thus,

\begin{equation}\label{15}
  \alpha\leq\frac{12 \times
  10^{-9}}{[\Psi(\mu,r_{s})-\Psi(\mu,r_{m})]}.
\end{equation}

According to (15), for a given $\lambda$ in the range $1.2 \times
10^{7} \rm\: m-3.8 \times 10^{8} \rm\: m$, the $\alpha$ parameter
of the $\alpha-\mu$ model is constrained as in Fig. 3. This new
result improves the constraint in the range $1.2 \times 10^{7}
\rm\: m-3.8 \times 10^{8} \rm\: m$ on Fig. 2 by about one order of
magnitude.


\section{Discussions and Outlook}

\noindent In this paper, we have used the satellite measurement of
earth gravity, the lunar orbiter measurement of lunar gravity, and
laser ranging measurement to constrain the intermediate-range
gravity from $\lambda=1.2 \times 10^{7} \rm\: m-3.8 \times 10^{8}
\rm\: m$. For the satellite measurement of earth gravity, the
geodesy mission GOCE (Gravity field and steady-state Ocean
Circulation Explorer)\cite{24} is planned to be launched in 2006.
The scientific objective is the mapping of the Earth's static
gravity field with a very high resolution and accuracy on a global
scale. GOCE is a drag-free mission, flown in a circular and
sun-synchronous orbit at an altitude between 240 km and 250 km. A
successful mission presupposed, GOCE would finally deliver the
Earth's gravity field with a resolution of about 70 km
half-wavelength and a global geoid with an accuracy of about 1 cm.
When this and other geodesy missions are implemented, the
$GM_{earth}$ will be determined more accurately.

SMART-1,\cite{25} the Europe's first lunar mission launched in
2003, is playing a vital role in developing cutting edge
technologies that will be a major part of the future of lunar and
planetary science. Japan plans to launch LUNAR-A mission in
2005,\cite{26} whose scientific objective is to explore the lunar
interior using seismometry and heat-flow measurements. However,
improvement in lunar gravity model is not a goal of SMART-1 and
LUNAR-A.

Japan is planning to launch SELENE (SELenological and ENgineering
Explorer) using H-2A rocket in 2006.\cite{27} The project will use
a main orbiting satellite and two subsatellites to obtain highly
accurate lunar gravity model. After that, it will be possible to
construct a global gravity model of the Moon using spherical
harmonics, and the value of $GM_{moon}$ will be improved.

Lunar laser ranging at CERGA is under renovation and
improvement.\cite{28} APOLLO program will use the 3.5 m telescope
at the Apache Point Observatory for lunar laser ranging.\cite{29}
These experiments will push Lunar Laser Ranging into
millimeter-accuracy range. As a result, constrains in this paper
will also be improved.

For small $\lambda$, analysis of satellite ranging data together
with GOCE will improve the constraints. For lager $\lambda$, we
look into ASTROD I and ASTROD missions which are currently under
pre-phase A study. ASTROD I and ASTROD are to implement deep-space
pulse and interferometric laser ranging in the solar system, to
test relativistic gravity and to increase the sensitivity of
solar, planetary and asteroid parameter determination by 1 to 5
order of magnitudes. This will improve constraints on $\alpha$ for
larger $\lambda$ significantly.\cite{30,31,32,33}

In this paper, we have used data from earth gravity, lunar
gravity, and lunar laser ranging separately to constrain the
strength of intermediate-range gravity. In the future, we will use
the $\alpha-\mu$ model as the underlying gravity theory for a
grand fit of all three kinds of data together, With increased
accuracy of gravity measurements and lunar laser ranging, we
expect the constraints on intermediate-ranging gravity to be
significantly improved.

\nonumsection{Acknowledgements} \noindent We thank the National
Natural Science Foundation (Grant no. 10343001, 10475114,
10273024, 10203005, 10227301) and the Foundation of Minor Planets
of Purple Mountain Observatory for financial support.

\nonumsection{References}

\end{document}